**Probable cause for the superconductor-like properties of alkane-wetted graphite and single-layer graphene above room temperature under ambient pressure**

Myung-Hwan Whangbo*,a,b,c

a Department of Chemistry, North Carolina State University, Raleigh, NC 27695-8204, USA

b Group SDeng, State Key Laboratory of Structural Chemistry, Fujian Institute of Research on the Structure of Matter (FJIRSM), Chinese Academy of Sciences (CAS), Fuzhou 350002, China

c State key Laboratory of Crystal Materials, Shandong University, Jinan 250100, China

**Abstract**

Recently Kawashima has reported that, when wetted with alkanes, several forms of graphite and single-layer graphene exhibit superconductor-like properties above room temperature under ambient pressure [*AIP Adv*. **2013**, *3*, 052132; arXiv:1612.05294; arXiv:1801.09376]. Under the assumption that these seemingly unlikely properties arise from the presence of paired electrons brought about by the alkane-wetting, we explored their implications to arrive at a probable mechanism for strong electron-pairing driven by Fermi surface nesting and acoustic phonon. This mechanism explains why alkane-wetting is essential for the graphene systems to become "superconductor-like" above room temperature and why the "$T_c$" of alkane-wetted pitch-based graphite fibers increases almost linearly from ∼363 to ∼504 K with increasing the molecular weight of alkane from heptane to hexadecane. It also provides a number of experimentally-verifiable predictions, the confirmation of which will provide a strong support for the superconductivity driven by Fermi surface nesting and acoustic phonon.

Ever since the discovery of the superconducting phenomenon in 1911,[1] it has been a holy grail to find materials that superconduct around room temperature under ambient pressure. The guiding principle for finding new superconducting materials has been the Bardeen-Cooper-Schrieffer (BCS) theory,[2] which showed that the charge carriers of a superconducting state are pairs of electrons rather than individual electrons, and the electron pairing arises from electron-phonon interactions. In this theory the superconducting transition temperature $T_c$ of a metal is related to the average phonon frequency $\langle\omega\rangle$ and the density of states $n(e_F)$ at the Fermi level $e_F$ of the metal as

$$T_c = \langle\omega\rangle \exp\left[-\frac{1}{n(e_F)V}\right]$$

where V is the pairing potential, with the term $n(e_F)V$ representing the electron-phonon coupling constant $\lambda$. This BCS equation shows that $T_c$ will increase with increasing $\langle\omega\rangle$ and/or $n(e_F)$. Over the years, the BCS theory was extended,[3] and the BCS equation was refined.[4,5] These developments made it possible to predict $T_c$ of any conventional superconductor on the basis of first principles electronic structure calculations. Using this computational approach, Duan et al.[6] predicted $T_c$ = 191 – 204 K for $H_2S$ under pressure of 200 GPa. This predicted transition was experimentally confirmed by Drozdov et al.,[7] who found $T_c$ = 203 K for $H_2S$ under 153 GPa. So far, ∼200 K is the highest $T_c$ among all known superconductors. However, it is still considerably below room temperature and requires an extremely high pressure to achieve. Thus, it remains a challenging dream to find a material that superconducts around room temperature under ambient pressure.

However, there is a strong possibility that such a dream material has already been around us. A Josephson-tunneling behavior was reported for the carbon layer of an aluminum-graphite-aluminum sandwich in 1974[8] (12 years prior to the discovery of the first cuprate superconductor[9]), superconducting-like magnetization hysteresis loops for HOPG in 2000,[10] and a granular superconducting behavior in graphite powders in 2012.[11] Recently, Kawashima reported that several forms of graphite (i.e., graphite flakes, HOPG, pitch-based graphite fibers, and single-layer graphene) exhibit persistent currents and a very small resistance when wetted with various saturated alkanes,[12-14] that alkane-wetted pitch-based graphite fibers increase their "$T_c$" almost linearly from ∼363 to ∼504 K as the molecular weight of the alkane is increased from heptane to hexadecane,[13] and that single-layer graphene shows a "Meissner-like" effect when wetted with alkane.[14] Since the wetting of graphite samples with alkane affects primarily their surfaces, Kawashima's observations amount essentially to the phenomenon of alkane-wetted single-layer graphene. The alkanes covering the graphene surface are present in liquid state, so the alkane wetting should form a layer of liquid alkane on the graphene surface. Both graphene and alkane are nonpolar, and the interaction between the two is of physisorption type. This makes one wonder how alkane-wetting can bring about such profound changes in the transport properties of graphene as found by Kawashima.

The transport properties of graphene wetted with alkane, reported by Kawashima, are difficult to explain unless they are caused by paired electrons. Thus, we persue the implications of Kawashima's observations under the assumption that alkane-wetting does generate paired electrons in graphene above room temperature, though seemingly unlikely it may seem. The key to finding the remarkable role of alkane wetting plays is to ask what physical property of graphene quickly changes by the wetting, because the transport properties of graphene quickly change on wetting. The only substantial interaction that the graphene layer can have with the liquid alkane layer covering its surface is "mechanical". The phonon dispersion curves of an isolated single-layer graphene[15] show that only the acoustic modes of vibrations, in which all atoms of a unit cell move in-phase, can be excited by the temperature of ~360 − ~500 K and that the out-of-plane acoustic (ZA) mode is considerably lower in energy at any given wave vector than the transverse acoustic (TA) and longitudinal acoustic (LA) modes.[15] (The LA and TA modes oscillate parallel and perpendicular to the C-C bonds of graphene, respectively.) Therefore, the vibrations of a graphene layer in the temperature range of ~360 − ~500 K are dominated by the ZA mode in which all atoms of a unit cell move in-phase out of the graphene plane. Thus, it is expected that the amplitude of the ZA mode becomes quickly reduced by alkane wetting. This implies that the "superconductor-like" behavior of graphene results from the damping of the ZA mode. The latter enhances the TA and LA modes due to the phenomenon of the spectral weight transfer,[16] so the electron pairing above room temperture is most probably driven by the in-plane acoustic vibrations.

Now we turn our attention to the electroniuc structure of graphene. Single-layer graphene has the electronic structure characterized by a Dirac cone located at **K** and its symmetry-equivalent points in k-space, with the Fermi level $e_F$ at the Dirac point (**Figure 1a**) so that $N(e_F) = 0$.[17] Then, the BCS equation predicts $T_c$ = 0 for graphene, implying the absence of paired electrons. As already pointed out, Kawashima's observations cannot be explained unless paired electrons exist in alkane-wetted graphene. This conceptual impasse, plus the effect of alkane wetting on the acoustic modes of graphene vibration discussed above, leads us to question if there exists a high-temperature electron-pairing mechanism, hidden in plain sight, that has escaped our notice. Therefore, we search for the electronic structure of graphene that can couple with the TA and LA modes to generate strongly coupled electron pairs capable of prevailing over the disrupting effect of the lattice vibrations present above room temperature. Here we show that this is most likely the case by uncovering a probable electron-pairing mechanism that explains Kawashima's observations on alkane-wetted graphene. Implications of this mechanism lead to a number of predictions on alkane-wetted graphene that can be readily verified by experiments.

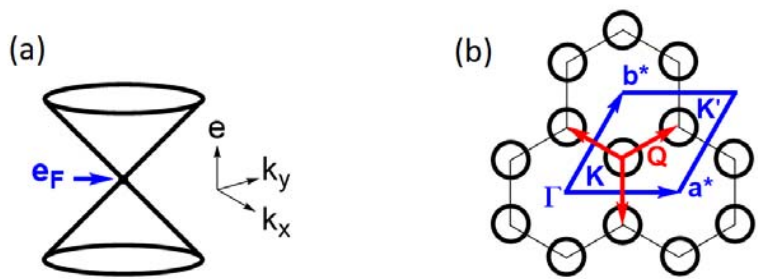

Figure 1. Electronc structure of single-layer graphene: (a) A perspective view of a Dirac cone located at **K**. (b) A top projection view of the Dirac cones in k-space, where each circle represents a Dirac cone.

Let us first consider a possible mechanism needed to support the electron pairing at high temperature. The BCS mechanism of electron pairing is essentially driven by the electron movement at low temperature as depicted in **Figure 2a**; an electron passing through the lattice drags nearby cations momentarily while another electron passing in the opposite direction follows the moving-away cations. Then, the two electrons appear effectively paired. In essence, the BCS electron pairing relies on small-energy-scale phonons, which become available at low temperature where large-energy-scale phonons are quenched. For superconductivity to occur around room temperature under ambient pressure, electron pairing should take place despite the presence of strong cation vibrations. An obvious way to fulfill this requirement is to utilize the available lattice vibrations for electron pairing. We need a paradigm shift in the way we think about how electron-pairing occurs. Instead of the electron movement dragging nearby cations as imagined in the BCS theory, one should consider the vibrating cations pulling passing-by electrons to form paired electrons. The frozen-phonon approximation posits that the relatively slow movement of the much heavier cations makes the electrons follow this movement almost instantaneously.[18] This approximation opens the way for a new microscopic mechanism of electron pairing driven by acoustic phonons, in which all lattice atoms of a unit cell oscillate in-phase. Consider the cations oscillating along a certain axis in an acoustic mode with wave vector $\mathbf{Q}$ and two electrons moving in the two opposite directions along the axis of the oscillation (**Figure 2b**). During the moment when the cations move in-phase in one direction, the two electrons will be both pulled toward the moving direction of the cations. Then, the two electrons would appear effectively as paired. This electron pairing will be strong at high temperature because the cations will have large vibrational displacements and implies that the paired electrons will possess an oscillation behavior with the frequency of the acoustic phonon driving the pairing.

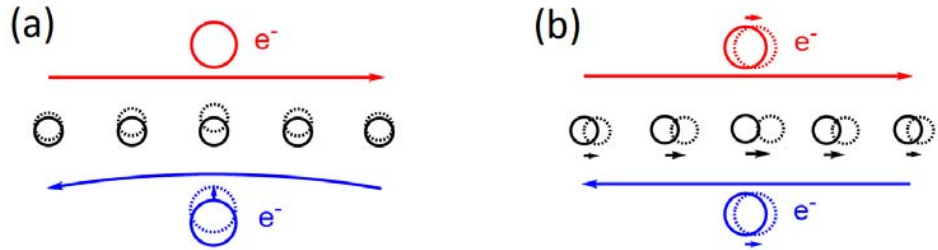

Figure 2. Schematic depiction of how effective electron pairing takes place in (a) the BCS emchanism at low temperature and (b) the mechanism driven by acoustic phonon at high temperature. The smaller circles represent the cations of the lattice, and the larger circles electrons.

To enable the electron pairing mechanism driven by the acoustic phonon of the lattice, the electronic structure of a metal should possess a certain feature. In general, a metal tends to lower its energy by opening a bandgap at the Fermi level typically in two ways. One is to form a charge density wave (CDW), which utilizes large-energy-scale phonons. The other is to form a superconducting state, which utilizes small-energy-scale phonons as already mentioned. Consider a metal with nested Fermi surface depicted in **Figure 3a,b**. The electronic states important for the formation of both CDW and superconducting states are those lying close to the Fermi level, namely, the states with their occupied wave vectors $\mathbf{k}_o$ and unoccupied wave vectors $\mathbf{k}_u$ lying close to the Fermi suface. A CDW state with propagation vector $\mathbf{Q}$ is formed by the interaction of the occupied states $\phi(\mathbf{k}_o\sigma)$ with the unoccupied states $\phi(\mathbf{k}_u\sigma)$ (here $\sigma$ represents either ↑ or ↓ spin) related by $\mathbf{k}_o \pm \mathbf{Q} = \mathbf{k}_u$ and $e(\mathbf{k}_o \pm \mathbf{Q}) = e(\mathbf{k}_u)$ (**Figure 3a**).[20-22] In contrast, the energy lowering that leads to a superconducting state in the BCS theory arises from the interaction of the occupied pair-states $\phi(\mathbf{k}_o \uparrow)\phi(-\mathbf{k}_o \downarrow)$ with the unoccupied pair-states $\phi(\mathbf{k}_u \uparrow)\phi(-\mathbf{k}_u \downarrow)$ (**Figure 3b**), where $\mathbf{k}_o \pm \mathbf{q} = \mathbf{k}_u$ and $e(\mathbf{k}_o \pm \mathbf{q}) = e(\mathbf{k}_u)$ with $\mathbf{q}$ representing a wave vector small in magnitude. The vector $\mathbf{q}$ needed to form a superconducting state is provided by the momentary displacements of the cations during the process of electron-phonon coupling, and the vector $\mathbf{Q}$ needed to form a CDW state by an acoustic mode of the lattice vibration. Consequently, $|\mathbf{q}/\mathbf{Q}| \ll 1$. Namely, the BCS electron pairing employs small-energy-scale phonons so that the associated energy lowering is not large, which makes the electron pairing weak. For a metal with nested surface to have strong electron pairing, the pair-state excitations[19] across the Fermi level must involve the large-energy-scale acoustic phonons of the nesting vector $\mathbf{Q}$. However, such a metal tends to form a CDW state,[20-22] because the formation of a CDW state provides a much stronger energy lowering than does that of a superconducting state of the BCS theory. If it were possible to remove the CDW instability of a metal with perfectly nested Fermi surface, it would be possible for the metal to utilize the large-energy-scale phonon associated with the nesting vector $\mathbf{Q}$ to substantially lower its energy by forming strongly coupled electron pairs. The CDW instability is weakened if the area of the nested Fermi surface is reduced. To remove this tendency completely while maintaining the state of perfect nesting, the Fermi surface area should be reduced to a single k-point. This is precisely the case for single-layer graphene.[17] At each Dirac cone, the Fermi surface is given by the Dirac point where the vertices of the two cones touch (**Figure 1a**), and adjacent Dirac points are completely nested by the vector $\mathbf{Q} = (\mathbf{a}^* + \mathbf{b}^*)/3$ and its equivalent ones (**Figure 1b**).

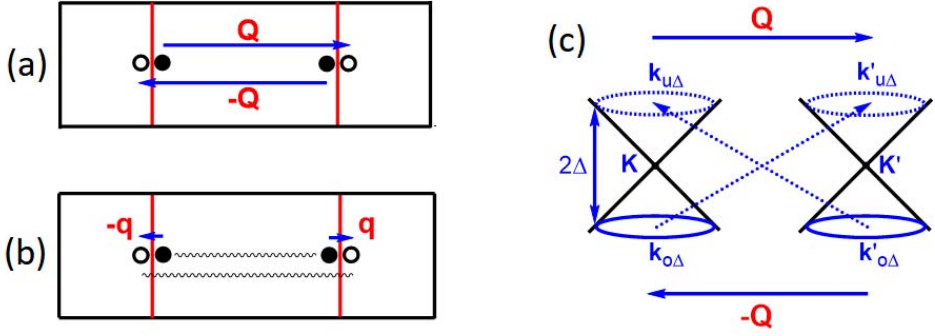

Figure 3. Illustrations of the excitations[19] (a) from $\phi(\mathbf{k}_o \sigma)$ to $\phi(\mathbf{k}_u \sigma)$ to form a CDW state, (b) from $\phi(\mathbf{k}_o \uparrow)\phi(-\mathbf{k}_o \downarrow)$ to $\phi(\mathbf{k}_u \uparrow)\phi(-\mathbf{k}_u \downarrow)$ to form a BCS superconducting state, and (c) from $\phi(\mathbf{k}_o \uparrow)\phi(-\mathbf{k}_o \downarrow)$ to $\phi(\mathbf{k}'_u \uparrow)\phi(-\mathbf{k}'_u \downarrow)$ as well as from $\phi(\mathbf{k}'_o \uparrow)\phi(-\mathbf{k}'_o \downarrow)$ to $\phi(\mathbf{k}_u \uparrow)\phi(-\mathbf{k}_u \downarrow)$ between adjacent Dirac cones to form a superconducting state in single-layer graphene by the FSNAP mechanism. In (a) and (b), the red lines represent the Fermi surfaces in two-dimensional representation. $\mathbf{k}_o \pm \mathbf{Q} = \mathbf{k}_u$ in (a), and $\mathbf{k}_o \pm \mathbf{q} = \mathbf{k}_u$ in (b). $\mathbf{k}_o \pm \mathbf{Q} = \mathbf{k}'_u$ and $\mathbf{k}'_o - \mathbf{Q} = \mathbf{k}_u$ in (c).

    As in the case of the BCS theory, we form pair-states using the k-points that lie within a certain energy cutoff, $\Delta_c$, from the Fermi level. (In the BCS theory, the energy cutoff $\delta_{BCS}$ is commonly taken to be the Debye temperature, which is 1813 K for an isolated graphene.[23] However, there is a report indicating that $\delta_{BCS}$ is only several times the $T_c$.[24]) In each Dirac cone (**Figure 3c**), we assume that the energy difference $2\Delta$ between the energy $e(\mathbf{k}_{o\Delta})$ on the occupied ring $\mathbf{k}_{o\Delta}$ and the energy $e(\mathbf{k}_{u\Delta})$ on the unoccupied ring $\mathbf{k}_{u\Delta}$ can be supplied by the available thermal energy when $\Delta \leq \Delta_c$. Consider now the two adjacent Dirac cones nested by the vector $\mathbf{Q}$ (**Figure 3c**). The occupied ring $\mathbf{k}_{o\Delta}$ of the Dirac cone at **K**

is nested by the vector $\mathbf{Q}$ to the unoccupied ring $\mathbf{k}'_{u\Delta}$ of the Dirac cone at $\mathbf{K}'$. Once shifted in k-space by $\mathbf{Q}$ (i.e., once collectively excited by the in-plane acoustic phonon $\mathbf{Q}$), the occupied ring $\mathbf{k}_{o\Delta}$ overlaps completely with the unoccupied ring $\mathbf{k}'_{u\Delta}$, namely, $\mathbf{k}_{o\Delta}+\mathbf{Q} = \mathbf{k}'_{u\Delta}$, but the energies e($\mathbf{k}_{o\Delta}+ \mathbf{Q}$) and e($\mathbf{k}'_{u\Delta}$) are different, i.e., e($\mathbf{k}_{o\Delta}+ \mathbf{Q}$)+2$\Delta$ = e($\mathbf{k}'_{u\Delta}$). Given the energy 2$\Delta$ supplied by the available thermal energy, the states on the shifted ring $\mathbf{k}_{o\Delta}+ \mathbf{Q}$ become equal in energy to those on the ring $\mathbf{k}'_{u\Delta}$, hence creating a "virtual Fermi surface". This occurs for all $\Delta$ values as long as $\Delta \leq \Delta_c$. Consequently, the occupied pair-states $\phi(\mathbf{k}_o \uparrow)\phi(-\mathbf{k}_o \downarrow)$ on the ring $\mathbf{k}_{o\Delta}$ at $\mathbf{K}$ can be excited to the unoccupied pair-states to $\phi(\mathbf{k}'_u \uparrow)\phi(-\mathbf{k}'_u \downarrow)$ on the ring $\mathbf{k}'_{u\Delta}$ at $\mathbf{K}'$ with the help of the in-plane acoustic phonon of wave vector $\mathbf{Q}$. For all possible values of $\Delta$ ($\leq \Delta_c$) the pair-state excitations employ the identical wave vector $\mathbf{Q}$, which represents a coherent, collective lattice vibration. Consequently, these pair-state excitations driven by the mechanism based on the Fermi surface nesting and acoustic phonon (FSNAP) could provide a strong enough electron pairing that can surmount the lattice vibrations present at high temperature. [The same conclusion is reached from the consideration of the excitations from the occupied pair-states $\phi(\mathbf{k}'_o \uparrow)\phi(-\mathbf{k}'_{ou} \downarrow)$ on the ring $\mathbf{k}'_{o\Delta}$ at $\mathbf{K}'$ to the unoccupied pair-states $\phi(\mathbf{k}_u \uparrow)\phi(-\mathbf{k}_u \downarrow)$ on the ring $\mathbf{k}_{u\Delta}$ at $\mathbf{K}$ by the vector $-\mathbf{Q}$.]

Let us now discuss the seemingly unlikely superconductor-like behaviors of alkane-wetted graphite and single-layer graphene from the viewpoint of the FSNAP mechanism. The nesting vector $\mathbf{Q}$ lies in the plane of graphene, so it is the in-plane acoustic modes of the graphene lattice that are necessary for the FSNAP pair-state excitations. The phonon dispersion curves of single-layer graphene show[15] that, for the ZA, TA and LA modes to reach the nesting vector $\mathbf{Q} = (\mathbf{a}^* + \mathbf{b}^*)/3$ along the $\Gamma \to K$ direction, the required temperatures are ~140, ~610 and ~850 K, respectively. At temperatures up to ~500 K, only the ZA mode can reach the $\mathbf{Q}$ value. However, the alkane-wetting will dampen the ZA mode, hence enhancing those of the TA and LA modes and consequently enabling the FSNAP electron pairing mechanism to take place. The "$T_c$" of alkane-wetted pitch-based graphite fibers, defined as the temperature at which the resistance increases sharply on raising temperature, increases almost linearly from ~363 to ~504 K with increasing the molecular weight of the alkane from heptane to hexadecane.[13] This result means that the ZA mode of graphene is more effectively weakened with increasing the molecular weight of alkane, which in return indicates that the primary role of alkane-wetting is to weaken the ZA phonon mode of graphene. The above discussion suggests that other molecules in liquid state such as alkanols can also be used to wet graphene and make the FSNAP pairing mechanism operate above room temperature. It is important to find organic molecules that will bring down the "$T_c$" around room temperature. The approximately-linear relationship[13] between "$T_c$" and the molecular weight of alkane suggests that wetting graphene with pentane (bp = 36.1 °C, ρ = 0.626 g/cm$^3$) might lead to "$T_c$" ≈ 300 K, but the high vapor pressure of pentane makes it impracticable. 1-pentanol (bp = 137 – 139 °C, ρ = 0.818 g/cm$^3$) or 2-pentanol (bp = 119.3 °C, ρ = 0.812 g/cm$^3$) might be reasonable alternatives.

Nonmagnetic superconductors known so far maintain their superconducting states at all temperatures below their $T_c$. The same is not expected for superconductors induced by the FSNAP electron pairing. Given the high temperature where this mechanism operates, the nesting between the $\mathbf{k}_{o\Delta}$ and $\mathbf{k}'_{u\Delta}$ rings (as well as that between the $\mathbf{k}_{u\Delta}$ and $\mathbf{k}'_{o\Delta}$ rings) can be approximately mainained by the acoustic phonons of wave vectors $\mathbf{Q} \pm \delta\mathbf{Q}$ (**Figure 4a**), where $\delta\mathbf{Q}$ represents a small deviation from the perfect nesting. Consequently, for the wave vectors between $\mathbf{Q}_b$ and $\mathbf{Q}_a$ satisfying the conditions $\mathbf{Q} - \delta\mathbf{Q} \leq \mathbf{Q}_b \leq \mathbf{Q}$ and $\mathbf{Q} \leq \mathbf{Q}_a \leq \mathbf{Q} + \delta\mathbf{Q}$, respectively, the superconductivity based on the FSNAP mechanism is expected. Above a certain temperature $T_c^a$, the maximum wave vectors $\mathbf{Q}_L$ of the TA and LA modes become greater than $\mathbf{Q}_a$ in magnitude (**Figure 4b**). Below a certain temperature $T_c^b$, the maximum wave vector $\mathbf{Q}_S$ of the TA and LA modes of graphene will be smaller than $\mathbf{Q}_b$ in magnitude. In both cases, the acoustic-phonon-mediated nesting cannot occur so that the FSNAP electron pairing mechanism does not operate. Thus, the superconductivity will disappear not only above $T_c^a$ but also below $T_c^b$. In short, the FSNAP electron pairing mechanism predicts the occurrence of superconductivity between two critical temperatures $T_c^b$ and $T_c^a$. So far, only the values of $T_c^a$ have been measured for pitch-based graphite fibers wetted with alkanes.[13] (The same consideration applies to the overlap between the $\mathbf{k}_{u\Delta}$ and $\mathbf{k}'_{o\Delta}$ rings.)

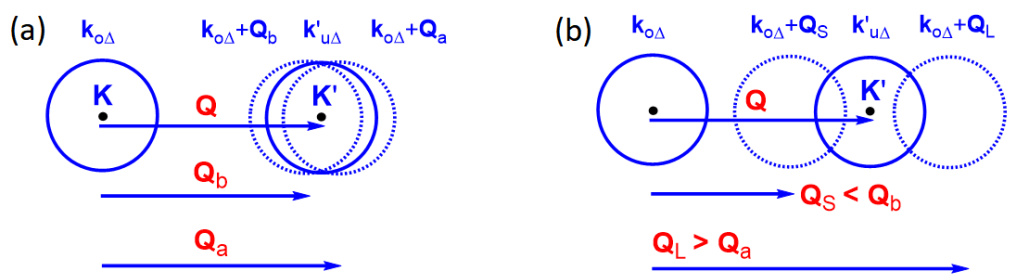

Figure 4. Illustration of why the superconductivity driven by the FSNAP mechanism of electron pairing leads to two critical temperatures.

In summary, if the "superconductor-like" behaviors of alkane-wetted graphene above room temperature were to originate from the presence of paired electrons, they should be strongly-coupled electron pairs so as to avoid being broken down at high temperature and hence must involve large-energy-scale phonons such as the in-plane acoustic phonons of graphene. The electronic structure of single-layer graphene has a perfectly nested Fermi surface with zero Fermi surface area and the nesting vector $\mathbf{Q} = (\mathbf{a}^* + \mathbf{b}^*)/3$. The latter leads to no CDW instability and hence allows a strong electron pairing to take place using the in-plane acoustic phonon of wave vector $\mathbf{Q}$. The electron pairing mechanism driven by the Fermi surface nesting and acoustic phonon provides natural explanations for the seemingly unlikely observations of Kawashima, and leads to a number of experimentally-verifiable predictions. The latter include the occurrence of superconductivity between two critical temperatures $T_c^b$ and $T_c^a$, an oscillation behavior of paired electrons reflecting the frequency of the acoustic phonon driving the pairing, and achieving superconductivity around room temperature by using other organic molecules in liquid state (e.g., 1-alkanol and 2-alkanol). Experimental verifications of these predictions will strongly support the occurrence of superconductivity driven by the Fermi surface nesting and acoustic phonon.

**Conflicts of Interest**

The author declares no conflict of interest.

**Acknowledgements**

8
The author would like to thank Shuiquan Deng and Jürgen Köhler, who have rekindled his interests in superconductivity.